\documentclass[amsmath,amssymb,prd, notitlepage,nofootinbib,twocolumn, superscriptaddress]{revtex4-2}

\usepackage{bm}
\usepackage{hyperref}
\usepackage{comment}

\newcommand{\aap}{{Astron.~Astrophys.}}
\newcommand{\jcap}{{JCAP}}
\newcommand{\apjl}{{Astrophys.~J.~Lett.}}
\newcommand{\apjs}{{Astrophys.~J.~Supp.}}
\newcommand{\aj}{{Astron.~J.}}
\newcommand{\mnras}{{Mon.~Not.~R.~Astron.~Soc.}}
\newcommand{\pasj}{{Publ.~Astron.~Soc.~Japan}}

\newcommand{\Planck}{{\it{Planck}}}

\usepackage{graphicx}
\usepackage[large]{subfigure}
\usepackage{amssymb, amsmath}
\usepackage[amssymb]{SIunits}
\usepackage{natbib}

\usepackage{xcolor}
\usepackage{booktabs}
\usepackage{multirow} 
\usepackage{caption}
\usepackage{subcaption}
 
\begin{document}

\preprint{APS/123-QED}

\title{Constraints on local primordial non-Gaussianity with 3d Velocity Reconstruction from the Kinetic Sunyaev-Zeldovich Effect}

\author{Alex Lagu\"e}%
\affiliation{Department of Physics and Astronomy, University of Pennsylvania, 209 South 33rd Street, Philadelphia, PA, USA 19104}%

\author{Mathew S. Madhavacheril}
\affiliation{Department of Physics and Astronomy, University of Pennsylvania, 209 South 33rd Street, Philadelphia, PA, USA 19104}%
\author{Kendrick M. Smith}
\affiliation{Perimeter Institute for Theoretical Physics, Waterloo, ON N2L 2Y5, Canada}
\author{Simone Ferraro}
\affiliation{Lawrence Berkeley National Laboratory, 1 Cyclotron Road, Berkeley, CA 94720, USA}
\affiliation{Berkeley Center for Cosmological Physics, University of California, Berkeley, 110 Sproul Hall 5800 Berkeley, CA 94720, USA}

\author{Emmanuel Schaan}
\affiliation{SLAC National Accelerator Laboratory, Menlo Park, CA 94305, USA}

\date{\today}
\smallskip

\begin{abstract}
The cosmic velocity field is an unbiased probe of the total matter distribution but is challenging to measure directly at intermediate and high redshifts. The large-scale velocity field imprints a signal in the cosmic microwave background (CMB) through the kinetic Sunyaev-Zeldovich (kSZ) effect. We perform the first 3d  reconstruction of the large-scale velocity field from the kSZ effect by applying a quadratic estimator to CMB temperature maps and the 3d positions of galaxies. We do so by combining CMB data from the fifth data release of the Atacama Cosmology Telescope (in combination with \Planck) and a spectroscopic galaxy sample from the Sloan Digital Sky Survey. We then measure the galaxy-velocity cross-power spectrum and detect the presence of the kSZ signal at a signal-to-noise ratio of 7.2$\sigma$. Using this galaxy-velocity cross-correlation alone, we constrain the amplitude of local primordial non-Gaussianity finding $f_{\rm NL}=-90^{+210}_{-350}$. This pathfinder measurement sets the stage for joint galaxy-CMB kSZ constraints to significantly enhance the $f_{\rm NL}$ information obtained from galaxy surveys through sample variance cancellation.
\end{abstract}
\maketitle

\section{Introduction}

Unraveling the initial conditions that seeded structure formation is one of the main objectives of modern cosmology. From a diverse set of observational datasets, we have determined that the early density field is well described by a Gaussian random field~\cite{Planck2015Non-Gaussian}. However, small deviations from pure Gaussianity, referred to as primordial non-Gaussianities (PNG), could yield crucial information about the physics of inflation (see e.g. \cite{1903.04409}). For this Letter, we will focus on local-type PNG quantified by the parameter $f_{\rm NL}$ through the relation~\cite{Komatsu2001AcousticSignatures}
\begin{align}
    \Phi = \phi + f_{\rm NL}\left(\phi^2-\langle \phi^2\rangle\right),
\end{align}
where $\Phi$ is the total gravitational potential and $\phi$ is the potential for a Gaussian field. The strongest constraints on local PNG have been obtained from the CMB from \textit{Planck} with $f_{\rm NL}=0.8\pm 5.0$~\cite{Planck2015Non-Gaussian} while large-scale structure probes have found $f_{\rm NL}=7\pm 31$ using the galaxy power spectrum and bispectrum~\cite{Damico2022LimitsOn},  $-4<f_{\rm NL}<27$ using the eBOSS quasar samples~\cite{Cagliari2023OptimalConstraints}, and $-87 < f_{\rm NL} < 19$ from cross-correlations between large-scale structure (LSS) tracers and CMB lensing~\cite{McCarthy2023ConstraintsOn}. 

The former CMB constraints probe non-Gaussianity through three-point functions in the temperature and polarization maps, while the latter LSS probes exploit the fact that a non-zero $f_{\rm NL}$ imprints a scale-dependent bias to dark matter tracers such as galaxies such that~\cite{Dalal2008ImprintsOf}
\begin{align}
    b_g^{\rm NL}(k) = b_g + f_{\rm NL}\frac{3 H_0^2 \Omega_m\delta_c (b_g -1)}{k^2T(k)D(z)}, \label{eq:bg_fnl}
\end{align}
where $k$ is the magnitude of the comoving Fourier wavenumber, $H_0$ is the Hubble constant, $\Omega_m$ is the relic matter density, $b_g$ is the galaxy bias, $T(k)$ is the matter transfer function, $\delta_c=1.42$ is the critical overdensity, and $D(z)$ is the linear growth factor at redshift $z$. Due to the $1/k^2$ scaling of the non-Gaussian bias, most of the information about $f_{\rm NL}$ is located on the largest scales. However, due to cosmic variance, few independent realizations of the large-scale modes exist. To overcome this sample variance, it is helpful to complement the galaxy power spectrum measurement (proportional to $\left(b_g^{\rm NL}\right)^2$) with information from a different tracer of the matter density \cite{astro-ph/0807.1770}.  As first suggested in \cite{Munchmeyer2019ConstrainingLocal}, the kinetic Sunyaev-Zeldovich (kSZ) effect in CMB maps can be used to reconstruct the velocity field, providing the necessary tracer that can dramatically improve constraints on $f_{\rm NL}$ through sample variance cancellation.  Recently, the kSZ effect was used to set bounds on $f_{\rm NL}$ using a two-dimensional analog of the approach we demonstrate here ~\cite{Krywonos2924ConstraintsOn}, albeit without a detected kSZ signal. A similar 2d kSZ estimator was also used with the sample of luminous red galaxies from the Dark Energy Spectroscopic Instrument (DESI) reaching a detection significance of $3.8\sigma$~\cite{McCarthy2024LargeScaleVelocity}. In this Letter, we constrain $f_{\rm NL}$ through its effect on scale-dependent bias using the full 3d information in the velocity field and detecting the kSZ signal at $>7\sigma$.

\begin{figure*}
    \centering
    \includegraphics[width=.8\linewidth]{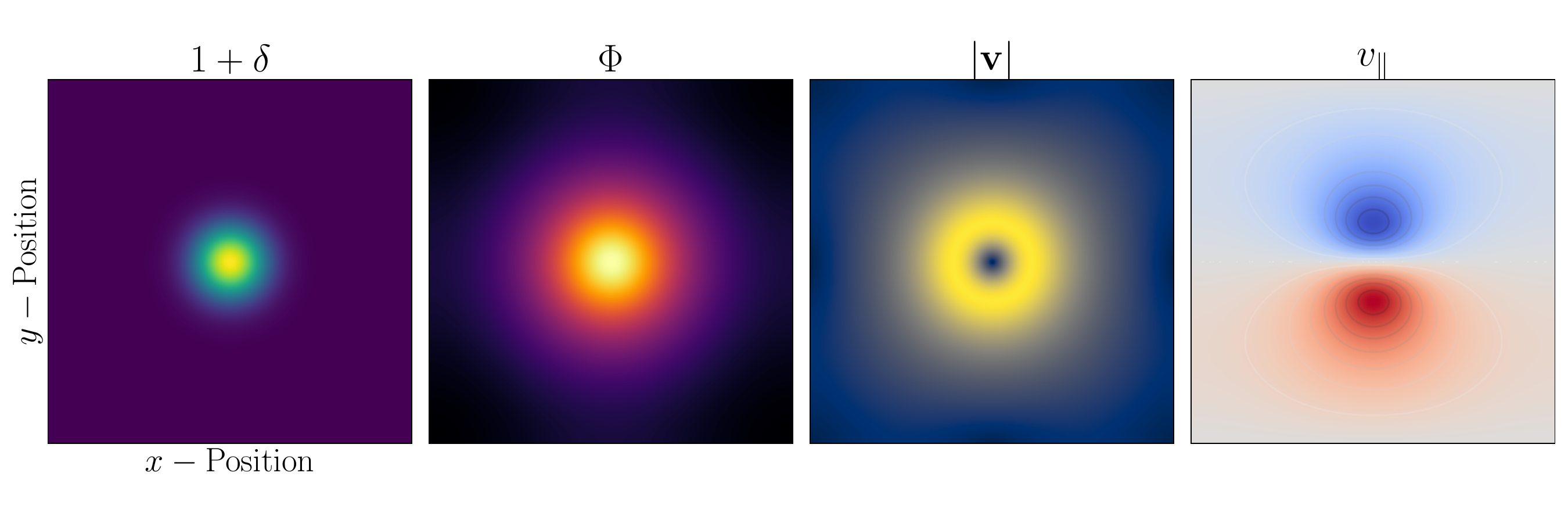}
    \caption{Sketch diagram of a two-dimensional Gaussian overdensity ($1+\delta$) and corresponding gravitational potential ($\Phi$), velocity magnitude ($|\mathbf{v}|$), and line-of-sight velocity ($v_\parallel$). We note the dipolar pattern which emerges when considering the parallel component of the velocity (the $y$-axis is taken as the line-of-sight). 
    }
    \label{fig:dipole-cartoon}
\end{figure*}

On large scales, the galaxy velocity field is an unbiased tracer of the total matter density, and the radial velocity field (along the line-of-sight) can be obtained from the continuity equation
\begin{align}
    v_r\left(\mathbf{k}\right)=  faH\frac{ik_r}{k^2}\delta_m\left(\mathbf{k}\right), \label{eq:vel-from-delta}
\end{align}
where $f$ is the logarithmic growth factor, $a=1/(1+z)$ is the scale factor, $H$ is the Hubble expansion rate, and $\mathbf{k},\;k_r,\;k$ are the comoving wavenumber, its radial component and its magnitude, respectively. The cosmic velocity field is thus a perfect complement to the galaxy density field if one hopes to get past the cosmic variance limit on large scales. The challenge in this case, however, is to measure said velocities. In one instance, we can use standard candles located within galaxies to make peculiar velocity measurements~\cite{Colless2001the2df}. This gives precise measurements but is limited to low redshifts of about $z\lesssim 0.15$~\cite{Saulder2023TargetSelection}. This reduces the volume of the survey and the range of scales we can probe, thus limiting the available information to constrain (local) primordial non-Gaussianity. We can also reconstruct the velocity field from the galaxy density~\cite{Schaan2016EvidenceFor,Schaan2021Combinedkinematic,Ried2024VelocityReconstruction,Hadzhiyska2024VelocityReconstruction}. However, this requires us to \emph{assume} the form of the galaxy bias and cannot be used to probe its scale-dependence. This type of velocity reconstruction is also quite sensitive to non-linear growth of structure and, therefore, is unsuitable to extract more information about the cosmic velocity field.

The kSZ effect is a CMB secondary which results from the motion of electrons relative to the rest frame of the CMB. As CMB photons travel through ionized gas, they inverse Compton scatter with free electrons and gain (or lose) energy proportional to the electron line-of-sight velocity ($v_{e, \;r}$)
\begin{align}
    \frac{\Delta T_{\rm kSZ}}{T_{\rm CMB}} =  \int  dz \frac{d\chi}{dz}  K(z) (1+\delta_{ e}) v_{e, \;r},
\end{align}
where $\chi$ is the comoving distance, $\delta_e$ is the electron density contrast. The kSZ prefactor reads
\begin{align}
    K(z) \equiv \sigma_T n_{ e, 0} (1+z)^2  x_{ e}(z) e^{-\tau(z)},
\end{align}
where $\sigma_T$ is the Compton cross-section, $n_{e, 0}$ is the present-day electron number density, $x_e$ is the ionization fraction, and $\tau$ is the optical depth.


The kSZ effect induces a squeezed bispectrum between long-wavelength ($k_L\sim 0.01\;\mathrm{Mpc}^{-1}$) radial velocities, short-wavelength ($k_S\sim 1\;\mathrm{Mpc}^{-1}$) galaxy overdensities, and CMB secondary temperature fluctuations~\cite{Smith2018KSZTomography}. From this bispectrum, one can construct a 3-dimensional quadratic estimator for the radial momentum field~\cite{Smith2018KSZTomography,Munchmeyer2019ConstrainingLocal,Giri2022ExploringKSZ}
\begin{widetext}
\begin{align}
    \hat{v}_r \left(\mathbf{k}_L\right) =  N_{v_r}^{(0)}\left(\mathbf{k}_L\right) \frac{K_{\rm eff}}{\chi^2_{\rm eff}} \int \frac{d^3k_S}{(2\pi)^3}\frac{d^2\ell}{(2\pi)^2} \frac{P_{ ge}(k_S)}{C_\ell^{\rm tot} P_{ gg}^{\rm tot}(k_S)}\delta_{ g}^*(\mathbf{k}_S) T_{\rm CMB}^*(\boldsymbol{\ell}) (2\pi)^3 \delta_D^3\left(\mathbf{k}_L+\mathbf{k}_S+\frac{\boldsymbol{\ell}}{\chi_\mathrm{eff}}\right),\label{eq:full-QE}
\end{align}
\end{widetext}
where $K_{\rm eff}, \; \chi_{\rm eff}$ are evaluated at $z_{\rm eff}$ and $\delta_D^3$ denotes the three-dimensional Dirac delta. The $P_{ gg}$, $P_{ ge}$, and $C_\ell^{\rm tot}$ denote the 3d galaxy-galaxy, 3d galaxy-electron, and CMB temperature map angular power spectra, respectively. Here, $T_{\rm CMB}(\boldsymbol{\ell})$ is the 2d Fourier transform of the beam-deconvolved CMB temperature map and $\delta_{ g}^*(\mathbf{k}_S) $ is the 3d Fourier transform of the galaxy overdensity field.   We can simplify the above by directly applying a filter to the CMB temperature map, giving us an approximated kSZ temperature map through
\begin{align}
    \hat{T}_{\rm kSZ}(\boldsymbol\theta) = \int \frac{d^2\ell }{(2\pi)^2} e^{i\boldsymbol\theta\cdot\boldsymbol\ell} \frac{P_{ ge}(\ell/\chi_{\rm eff})}{C_\ell^{\rm tot} P_{ gg}^{\rm tot}(\ell/\chi_{\rm eff})}T_{\rm CMB}(\boldsymbol\ell).
    \label{eq:Tmap}
\end{align}
The galaxy auto-spectrum and the galaxy-electron cross-spectrum are obtained using the halo model prescription implemented in the \texttt{hmvec} code~\footnote{\url{https://github.com/simonsobs/hmvec}}\cite{Smith2018KSZTomography}. The reconstruction in Eq. \ref{eq:full-QE} produces a ``box'' of the 3d momentum field, whose noise per mode on large scales tends to a constant
\begin{align}
    N_{v_r}^{(0)} = \frac{\chi^2_{\rm eff}}{K^2_{\rm eff}} \left(\int \frac{k_S dk_S}{2\pi}\; \frac{\left[ P_{ge}\left(k_S\right)\right]^2}{P_{gg}^{\rm tot}\left(k_S\right)C_\ell^{\rm tot}}\right)^{-1},\label{eq:Nvr}
\end{align}
where the CMB power spectrum is evaluated at $\ell=k_S/\chi_{\rm eff}$. With the noise being constant, we can calculate a temperature-based velocity weight by rescaling the approximated kSZ map through
\begin{align}
    \hat{T}_{v_r}(\boldsymbol{\theta}, z) = N_{v_r}^{(0)}\frac{K_{\rm eff}}{\chi^2_{\rm eff}} \hat{T}_{\rm kSZ}(\boldsymbol{\theta}). \label{eq:main-vr-recon}
\end{align}

Since the exact distribution of electrons on cosmological scales is unknown, our choice of model for the galaxy-electron cross-spectrum introduces a bias in the reconstructed velocity field such that~\cite{Smith2018KSZTomography}
\begin{align}
    \langle \hat{v}_r(\mathbf{k}) \rangle = b_v(\mathbf{k}) v_r(\mathbf{k})
\end{align}
where $b_v$ is the velocity reconstruction bias. It has been shown to be constant on large scales~\cite{Giri2022ExploringKSZ}. This bias stems from the optical depth degeneracy of kSZ measurements and the uncertainties of baryonic physics. If our fiducial $P_{ge}$ captures the true galaxy-electron correlation, we expect this bias to converge to unity. Since the true $P_{ge}$ is not known a priori, we focus on large scales and marginalize over this parameter in our analysis. Since the reconstruction bias is perfectly degenerate with the combination $f\sigma_8$, our estimator cannot measure the growth function as with peculiar velocities surveys~\cite{Qin2019TheRedshift}. This degeneracy can be broken in a sufficiently high SNR measurements by modeling the galaxy-velocity octupole in combination with the dipole.
\begin{figure}
    \centering
    \includegraphics[width=1.0\linewidth]{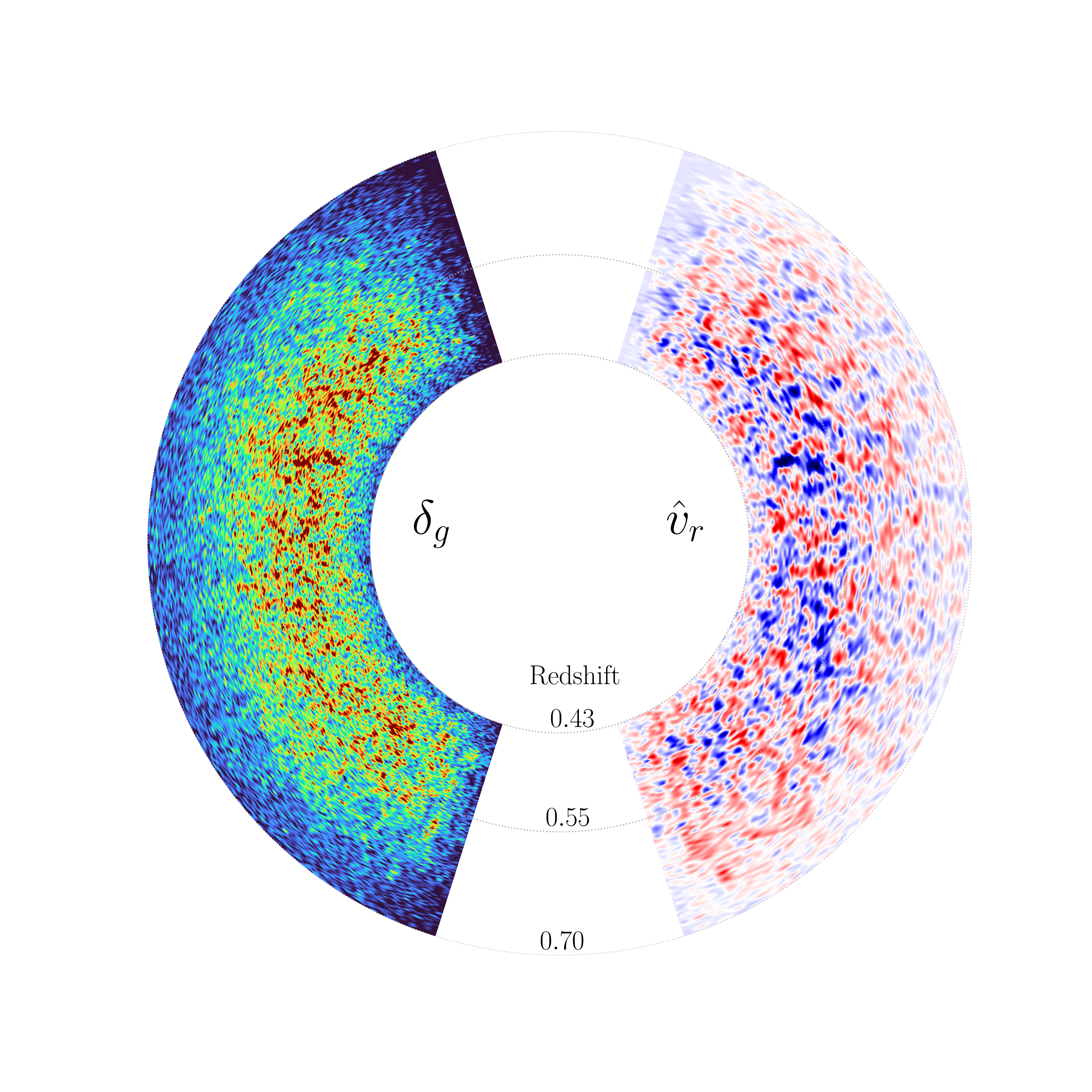}
    \caption{\textit{(Left panel}) Galaxy density in the CMASS NGC sample used in the reconstruction process. Darker shades correspond to higher density regions. (\textit{Right panel}) Smoothed map of the reconstructed line-of-sight momentum using the quadratic estimator on the CMASS NGC sample. The angular direction denotes right ascension and the field are summed along the declination axis. The bluer regions denote motion towards the observer in the velocity panel. 
    }
    \label{fig:gal-vel-map}
\end{figure}

\subsection{Data}
To measure the kSZ signal, we use CMB data from the Atacama Cosmology Telescope (ACT) and the \Planck~ satellite together with galaxy positions and spectroscopic redshifts from the Sloan Digital Sky Survey Baryon Oscillation Spectroscopic Survey twelfth data release (SDSS BOSS DR12; \cite{1208.0022,1607.03155}).  The CMB maps we use are ACT+\Planck~ co-adds \cite{2007.07290} at central frequencies of roughly 90 and 150 GHz (labeled \texttt{f090} and \texttt{f150}). We describe the processing and filtering of these maps in Appendix~\ref{app:map-prep}. For our galaxy sample, we limit our analysis to the North Galactic Cap (NGC) of the CMASS survey with $0.43<z<0.7$ and $z_{\rm eff}=0.55$, since it it is the largest contiguous area from BOSS that has overlap with ACT data. We also do not include the LOWZ sample as its low effective redshift does not allow us to reconstruct sufficiently large scales.

\section{Methodology}

\subsection{Galaxy-Velocity Cross-Correlation}

To calculate the correlation between the galaxy and velocity field, we use an adapted version of the Yamamoto estimator for the galaxy auto-spectrum~\cite{Yamamoto2006AMeasurement}. More specifically, we combine the latter with the peculiar momentum power spectrum estimator as developed in Ref.~\cite{Howlett2019TheRedshift}. We start with the weighted fields
\begin{align}
    F^g(\mathbf{x}) &= \frac{w_g(\mathbf{x}) [n_g(\mathbf{x}) - \alpha n_s(\mathbf{x})] }{A_g}\\
    F^v(\mathbf{x}) &= \frac{w_v(\mathbf{x}) n_g(\mathbf{x}) \hat{T}_{v_r}(\mathbf{x})}{A_v} \label{eq:F_v}
\end{align}
where $w_{g, v}$ are the weights, $A_{g, v}$ are the normalization terms, $\hat{T}_{v_r}$ is the velocity weight extracted from the temperature map, and $n_s$ is the uniform random catalog with a number of objects $1/\alpha$ times the number of galaxies in the survey. What distinguishes our reconstruction method from those of similar works is that Eq.~(\ref{eq:F_v}) describes a \textit{three-dimensional} velocity field. This is the first reconstruction using data of this kind. For our kSZ velocity reconstruction, we use as our velocity measurements the filtered and normalized temperature estimates of Eq.~(\ref{eq:main-vr-recon}). The velocity field estimator in Eq.~(\ref{eq:F_v}) scales as $F_v \sim \delta_g T$, and is our implementation of $\hat{v}_r$ from Eq.~(\ref{eq:full-QE}) for galaxy surveys. The normalization terms are given by
\begin{align}
    A_{g,v} = \int d^3x \;\bar{n}(\mathbf{x}) w_{g,v}(\mathbf{x}).
\end{align}
for the galaxies, we use the well-known FKP weights~\cite{Feldman1994Power-Spectrum} $w_g^{-1} = 1+\bar{n}(z)P^g_0$. For the velocity field we use the completeness weights. The momentum weights of Ref.~\cite{Howlett2019TheRedshift} are set to $w_v^{-1} = \langle v^2(z) \rangle+\bar{n}(z)P^v_0$ where $\langle v^2(z) \rangle$ is the variance in the velocity field as a function of redshift. In both cases, we also use the completeness weights. For our large-scale estimator, the real-space velocity field is far noisier than the momentum field measured with peculiar velocity surveys. The minimal variance weights chosen in that instance may not necessarily be optimal to extract a non-Gaussian signal~\cite{Cagliari2023OptimalConstraints}. We use $\langle v^2 \rangle = 1500^2\;\mathrm{km}^2/\mathrm{s}^2$ and $P_0^{v} = 5\times 10^{9} \;(\mathrm{Mpc}/h)^3(\mathrm{km}/\mathrm{s})$. The higher $P_0^{v}$ puts more weight on large-scales (higher redshifts) than peculiar velocity surveys~\cite{Howlett2019TheRedshift}, and helps reduce uncertainty of $f_{\rm NL}$ when studying non-Gaussianity. The approximate velocity variance is obtained by integrating the fiducial velocity power spectrum
\begin{align}
    \langle v^2 \rangle \approx \int \frac{k^2 dk}{6\pi^2} \left( \frac{faH}{k} \right)^2 P_{mm}^{\rm fid}(k),
\end{align}
where $P_{mm}^{\rm fid}$ is the fiducial (linear) matter power spectrum evaluated at the effective redshift of the survey.

In the plane-parallel approximation, the power spectrum estimator reads
\begin{align}
    |\hat{P}_\ell^{gv}(k)| &= (2\ell +1)\int \frac{d\Omega_k}{4\pi} F^{g}(\mathbf{k})F_\ell^{v}(\mathbf{k})^\star \\&= (2\ell+1)\int \frac{d\Omega_k}{4\pi} \int d^3 x \int d^3x^\prime F^{g}(\mathbf{x}) \nonumber \\&\;\;\;\;\;\;\;\;\; \times F^{v}(\mathbf{x}^\prime)\mathcal{P}_\ell\left(\hat{\mathbf{k}}\cdot \hat{\mathbf{x}}^\prime\right) e^{i\mathbf{k}\cdot(\mathbf{x}-\mathbf{x}^\prime)},
\end{align}
where $\mathcal{P}_\ell$ is the Legendre polynomial of order $\ell$. The plane-parallel approximation makes the estimation of the power spectrum computationally possible but introduces an error at wide angles~\cite{Beutler2019InterpretingMeasurements}. We account for this effect along with the survey geometry in our model which we describe in the following section.

\subsection{Model}
For the purposes of this first demonstration of 3d velocity reconstruction for primordial non-Gaussianity constraints, we only use the galaxy-velocity correlation $P_{gv}(k)$.\footnote{As described in \cite{Smith2018KSZTomography}, when applied to the same data, this cross-correlation contains the same kSZ information as  various other  detections in the literature, including from the pairwise method \cite{1203.4219,1603.03904,1607.02139,2101.08374,2207.11937} and the template method \cite{1510.06442,Schaan2021Combinedkinematic}.}  The full power of this technique can be unlocked by combining with $P_{gg}(k)$ and $P_{vv}(k)$, or writing a likelihood at the field-level (as done on simulations in Ref.~\cite{Giri2022ExploringKSZ}), but this is beyond the scope of this Letter. Here, we specifically model the galaxy-velocity power spectrum dipole. The dipole nature of the signal is due to the fact that we are sensitive only to the radial part of the velocity field as shown schematically in Fig.~\ref{fig:dipole-cartoon}. This plot shows a two-dimensional Gaussian overdensity along with its gravitational potential found using Poisson's equation. The magnitude and radial components of the velocity are found using Eq.~(\ref{eq:vel-from-delta}) and by taking the $y$-axis as the line-of-sight direction. The parameters we vary are the velocity bias and, in some cases, $f_{\rm NL}$. The theoretical galaxy-velocity cross-spectrum dipole reads
\begin{align}
    P_{\ell=1}^{gv}(k) = b_v \left(b_g + \frac{3f}{5}\right)   \left(\frac{ifaH}{k}\right)P_{mm}^{\rm fid}(k). \label{eq:gv-model}
\end{align}
In the presence of primordial non-Gaussianty, this model inherits one more free parameter as we substitute $b_g\to b_g^{\rm NL}(k, f_{\rm{NL}})$ using the scale-dependent bias from Eq.~(\ref{eq:bg_fnl}).

\section{Results}
We confirm that we detect the kSZ signal from the combination of CMASS and ACT data. We find that our detection significance matches that of other kSZ-based methods (e.g. \cite{Schaan2021Combinedkinematic,2101.08374}) as we would expect given their equivalence as shown in Ref.~\cite{Smith2018KSZTomography}. The resulting momentum field is shown schematically in Fig.~\ref{fig:gal-vel-map} along with the projected galaxy density field from which it is obtained and the detection significance can be found in Table~\ref{tab:results}.

\begin{figure}
    \centering
    \includegraphics[width=1\linewidth]{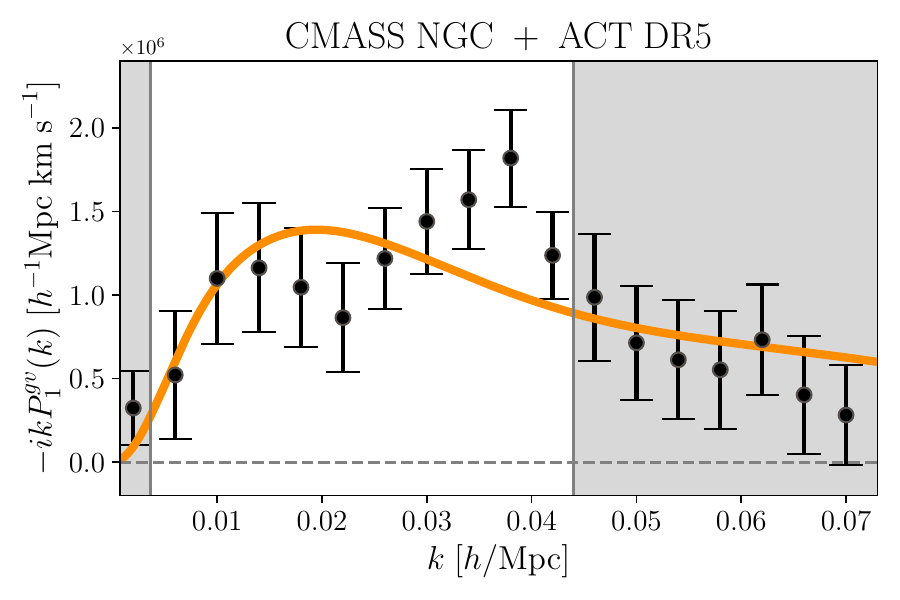}
    \caption{Galaxy-velocity dipole obtained with an inverse-variance combination of the $P_{gv}$ from the f090 and f150 maps. The solid orange line traces the best-fit model. The vertical grey shaded area denote the scales which are not used in our analysis.}
    \label{fig:combined-Pgv-data}
\end{figure}

We fix the parameters $\Omega_m=0.31$, $f(z_{\rm eff})=0.762$, $b_g=1.92$, and $H_0=67.37$ km/s/Mpc.  We compute the linear power spectrum using the Boltzmann code CLASS~\cite{Blas2011TheCosmic} using the \Planck\; 2018 best-fit cosmology~\cite{Planck2018}. We use the usual $\chi^2$ measure in a Gaussian likelihood which is given by
\begin{align}
    \chi^2 &= [\mathbf{t}(\Theta)-\mathbf{d}]^\mathrm{T} C^{-1} [\mathbf{t}(\Theta)-\mathbf{d}], \label{eq:chi2}
\end{align}
where $C$ is the covariance matrix, $\mathbf{t}$ is our theory prediction for the galaxy-velocity power spectrum given our parameters $\Theta$ which is either $\{b_v\}$ or $\{b_v, f_{\rm NL}\}$, and $\mathbf{d}$ is our data vector. For the latter, we combine the galaxy-velocity spectra from both CMB maps (\texttt{f090} and \texttt{f150}) by an inverse variance weighting. The covariance matrix is obtained from the CMB map itself and simulations as described in Appendix~\ref{app:cov-mat}. We display the final data points in Fig~\ref{fig:combined-Pgv-data}. We fit the resulting galaxy-velocity dipole to the model described in Eq.~(\ref{eq:gv-model}) . We estimate the signal-to-noise (SNR) as the square root of the difference in $\chi^2$ between the null hypothesis ($b_v=0$) and our best-fit. We add both frequency channels using a minimum variance combination of their power spectra, and we rule out the null hypothesis at $7.2\sigma$ with this result. We display the SNR for each frequency and their combination in Table~\ref{tab:results}. Our velocity bias is close to unity at $b_v=1.04$, suggesting that our choice of fiducial galaxy-electron power spectrum is close to the true underlying electron distribution on small scales. We also 
estimate the number of sigmas by numerically evaluating the Gaussian likelihood over an array of parameter values (which is trivial due to the fact that we have a single free parameter). As a validation step, we compute the cross-correlation between the kSZ-reconstructed velocities and the ones estimated from the continuity equation in Appendix~\ref{app:bao-recon-corr}.

\begin{table}[]
    \centering
    \begin{tabular}{c | c | c | c  }
        Frequency & Best-fit $b_v$ & SNR & $\chi^2$/dof\\\hline
        f090 & $0.842$ & $3.6\sigma$ & $17.7/(10-1)$  \\
        f150 & $1.17$ & $5.7\sigma$ & $5.67/(10-1)$ \\
        Combination & $1.04$ & $\mathbf{7.2\sigma}$ & $12.9/(10-1)$  \\
    \end{tabular}
    \caption{Signal-to-noise and goodness of fit measures for the two frequency bands and their combination for the NGC galaxy sample.}
    \label{tab:results}
\end{table}

\begin{figure}
    \centering
\includegraphics[width=0.7\linewidth]{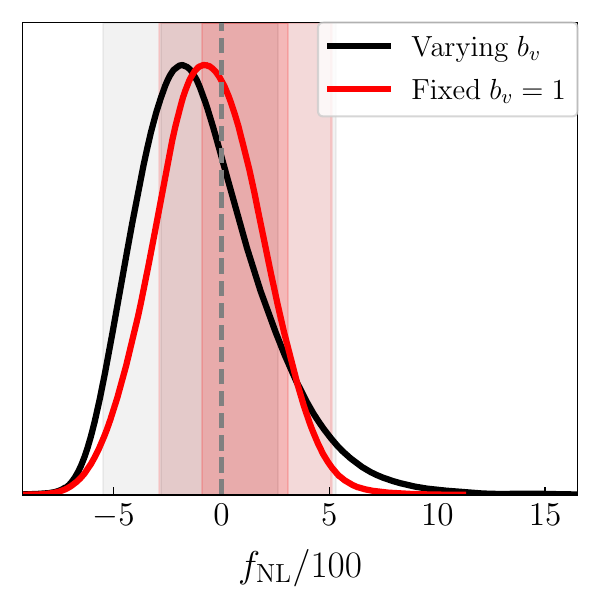}
    \caption{Posterior distribution for the local primordial non-Gaussianity parameter $f_{\rm NL}$ obtained from the reconstructed $P_{gv}$.}
    \label{fig:fnl-constraints}
\end{figure}
As part of our analysis, we also derive constraints on local primordial non-Gaussianity. We use the model described in Eq.~\ref{eq:gv-model} along with the modified galaxy bias $b^{\rm NL}_g(k)$. We vary $b_v$ and $f_{\rm NL}$ assuming a Gaussian likelihood as before. We use the Markov chain Monte Carlo sampling software \texttt{Zeus}~\cite{karamanis2020ensemble,karamanis2021zeus} to run the chains and the \texttt{GetDist}~\cite{Lewis:2019xzd} package for the postprocessing.

Marginalizing over $b_v$, we obtain the constraint $f_{\rm NL}=-90^{+210}_{-350}$ at the 68\% confidence level. We find some  degeneracy between the velocity bias and the level of non-Gaussianity, which degrades our constraints. This is due to the small range of wavenumbers used in the analysis. We limit our analysis on large-scales to avoid biases due to modeling errors due to wide-angle effects (see Appendix~\ref{app:systematics} for more details). For wavenumbers above $k\sim 0.03\;h/$Mpc, the velocity bias can become strongly scale-dependent, and we include only scales up to $k=0.043\;h/$Mpc. Accurately modeling the reconstructed $P_{gv}$ over a wider range of scales (both larger and smaller) would allow us to distinguish the scale-dependent bias from the velocity bias. We also fix the velocity bias to unity as an exercise and find $f_{\rm NL}=-60\pm 240$. Given our restricted sample and detection significance, these constraints on primordial non-Gaussianity are slightly weaker than the expected bounds. Using the galaxy-galaxy power spectrum monopole with the same data and $k$-range, we find an error $\sigma(f_{\rm NL})\approx 70$ (when keeping the galaxy bias fixed). From early Fisher forecasts, we expect the error to degrade by a factor of two when using the $P_{gv}$ compared to the $P_{gg}$~\cite{Munchmeyer2019ConstrainingLocal}. The uncertainty on $f_{\rm NL}$ is thus 70\% higher than this prediction. We attribute this to a combination of the wide-angle window function, the restricted $k$-range due to wide-angle effects, and the $N^{(3/2)}_{v_r}$ bias that was identified in Ref.~\cite{Giri2022ExploringKSZ}. This additional noise, labeled $N_{v_r}^{(3/2)}$, is proportional to $\langle v^2\rangle/\bar{n}$. This implies that this term will become subdominant in surveys with a much higher number density of tracers. We thus expect these constraints to improve significantly with lower CMB noise and larger galaxy samples.

\section{Conclusion}
In this Letter, we have presented the first 3d kSZ velocity reconstruction. This reconstruction is made using joint ACT+\Planck~ CMB data and the CMASS spectroscopic galaxy sample. We have computed the Fourier-space correlation between the velocities obtained using our algorithm and the galaxies of the CMASS NGC catalog and found our signal to be significant at over $7 \sigma$. This matches the detection significance of other kSZ-based analyses on similar datasets~\cite{Schaan2021Combinedkinematic}. Using an adapted model for the galaxy-velocity dipole, we have also constrained the amplitude of local non-Gaussian fluctuations in the early universe. The constraints we derived were weaker than expected from forecasts by a factor of about 1.7. We plan to investigate this in future work -- possible causes include the wide-angle window function, the restricted $k$-range due to wide-angle effects, and the $N^{(3/2)}_{v_r}$ bias from~\cite{Giri2022ExploringKSZ}.
Sources of noise beyond the $N_{v_r}^{(0)}$ of Eq.~(\ref{eq:Nvr}) are expected to decrease significantly when using surveys with a higher number density of tracers. 

Our measurements are the first of their kind and use the entire three-dimensional galaxy density field from spectroscopic redshift measurements. They allow for novel data combinations between the CMB and large-scale structure surveys. In particular, future measurements will use the galaxy-galaxy, galaxy-velocity, and velocity-velocity power spectra to overcome large-scale cosmic variance. We expect this technique to surpass constraints obtained from galaxy surveys alone in the near future~\cite{Munchmeyer2019ConstrainingLocal}. It was also recently shown in \cite{2407.21094} that constraints on beyond-local-type primordial non-Gaussianity \cite{1204.6324,1612.06366,2311.04882} can also be significantly improved (over using galaxy clustering alone) using the 3d kSZ tomography technique demonstrated in this work.  This measurement therefore opens a new window towards a wide variety of constraints on the physics of inflation using joint analyses of the CMB and large-scale structure.

\section*{Acknowledgments}
We thank Bruce Partridge, Fiona McCarthy, Florian Beutler, and Blake Sherwin for useful discussions. AL and MM acknowledge support from NASA grant 21-ATP21-0145.

\appendix

\section{Preparing the CMB Maps\label{app:map-prep}}

We use the co-added \Planck+ACT CMB maps from the DR5 release \cite{2007.07290}. We run our kSZ estimator separately on the 90 GHz and 150 GHz maps and on a difference of these maps, as a null test. We prepare a mask that allows 70\% of the full-sky based on Galactic dust emission seen by \Planck. We additionally mask regions where the 90 and 150 GHz maps have an RMS noise exceeding 70 $\mu K$-arcmin. After masking and transforming to spherical harmonics, we estimate empirical power spectra of the maps (pseudo-$C_{\ell}$ correcting for $f_{\rm sky}$) and fit a model $A C^{TT}_{\ell}b_{\ell}^2 + w$ with parameters $\{ A,w\}$, a fiducial lensed CMB temperature auto-spectrum $C^{TT}_{\ell}$ and known beam transforms $b_{\ell}$ in order to estimate the white noise level $w$. We compute the fiducial angular CMB temperature power spectrum  $C^{TT}_{\ell}$ with the Boltzmann code \texttt{CAMB}~\cite{Lewis:1999bs}, assuming the fiducial \Planck~ cosmology. These are used in constructing our CMB filter: we apply  $b_{\ell} / (C^{TT}_{\ell}b_{\ell}^2 + w)$ to the spherical harmonics of each frequency map to deconvolve the beam and isotropically inverse-variance filter it, representing the $C_{\ell}$ in the denominator of Eq.~(\ref{eq:Tmap}). We also addditionally filter by $P_{ge}/P_{gg}$ as shown in Eq.~(\ref{eq:Tmap}) to approximately apply the electron profile filter of Eq. \ref{eq:full-QE}.

\section{Correlation with BAO reconstruction\label{app:bao-recon-corr}}

\begin{figure}
    \centering
    \includegraphics[width=0.9\linewidth]{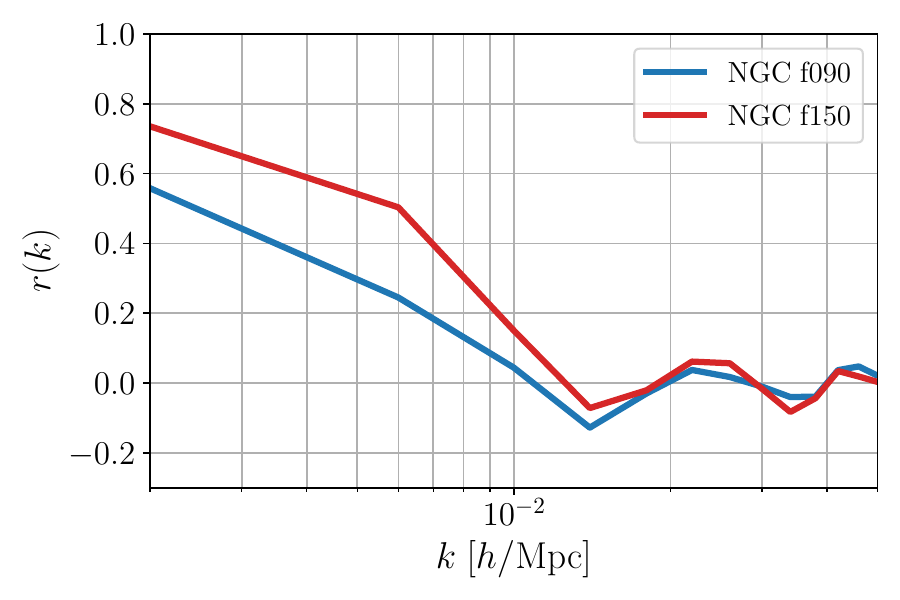}
    \caption{Approximate correlation coefficient between the BAO-reconstructed velocities and the kSZ velocities obtained from the quadratic estimator.}
    \label{fig:r-of-k}
\end{figure}

To test the quality of our reconstruction, we also estimate the correlation between the true and kSZ-reconstructed velocities. Since we do not have measurements of the galaxy peculiar velocities with CMASS, we employ a different reconstruction technique based on the linearized continuity equation applied to the galaxy density field. 
We follow the BAO reconstruction procedure implemented in the \texttt{pyrecon}~\footnote{https://github.com/cosmodesi/pyrecon} which solves the following equation for the galaxy velocity field~\cite{Nusser1994OnThe}
\begin{align}
    \nabla \cdot \mathbf{v} + f\nabla ( \mathbf{v} \cdot \hat{\mathbf{r}}) \hat{\mathbf{r}} = -faH\frac{\delta_g}{b_g},\label{eq:bao-recon}
\end{align}
where $\hat{\mathbf{r}}$ is the line-of-sight unit vector. We use the multi-grid reconstruction method as implemented in Ref.~\cite{Eisenstein2007ImprovingCosmological}. We henceforth refer to the velocities obtained by solving Eq.~(\ref{eq:bao-recon}) as \textit{BAO-reconstructed}. We measure the degree of correlation by the ratio of the cross and auto-spectra
\begin{align}
    r(k) \equiv \frac{P^{\hat v_r \bar v_r}_0(k)}{\sqrt{P^{\hat v_r \hat v_r}_0(k) P^{\bar v_r \bar v_r}_0(k)}},
\end{align}
where the bar and hat denote BAO and kSZ-based reconstruction, respectively. This correlation coefficient allows us to determine the maximum wavenumber ($k_{\rm max}$) for our analysis. As shown in Fig.~\ref{fig:r-of-k}, we find less than 10\% correlation on scales smaller than $k=0.03\;h/$Mpc. The correlation with BAO-reconstructed velocities drops quickly for wavenumbers $k>0.01\;h/$Mpc, but indicates that we are still signal-dominated in near $k\sim 0.01\;h/$Mpc. This matches the expectation from the $N$-body simulations of Ref.~\cite{Giri2022ExploringKSZ} although we do not find a perfect correlation on the largest scales. This lower correlation may be due to the BAO-reconstruction being noisy as well. As such, we are not directly correlating with the true halo velocities, as was done in Ref.~\cite{Giri2022ExploringKSZ}.

\section{Covariance Matrix\label{app:cov-mat}}
\label{sec:covapp}
To estimate the covariance of our measurement, we combine the variance of the noise and that of the signal under the assumption that they are independent. First, we extract the noise part of our covariance by running our quadratic estimator on the data map and a set of $N_\mathrm{s}=99$ simulated QPM CMASS mock catalogs~\cite{Lin2020TheCompleted} (this step also serves as a null test). The simulated galaxy catalogs have the same footprint and redshift distribution as the true CMASS data, they are uncorrelated with the data itself, and by extension the CMB map. In order to get an account for the variance in the signal, we create a set of simulated maps which include only the kSZ signal from the QPM catalogs. First, we perform a BAO-type reconstruction on the positions of the galaxies using the fiducial growth rate and galaxy bias. Then, we estimate the line-of-sight peculiar velocity of each galaxy from their displacements (from difference between pre and post-reconstructed positions). Finally, we use the \texttt{astropaint}\footnote{\url{https://github.com/syasini/AstroPaint}} algorithm to paint the kSZ signal to the sky and combine it to the simulated primary CMB. This step is done by assuming a Battaglia-type electron profile~\cite{Battaglia2016TheTau} and a constant halo mass of $3\times 10^{13}\;M_\odot$. We run our quadratic estimator on the correlated pairs simulated kSZ maps and QPM catalogs. The covariance is estimated as
\begin{align}
    C_{ij} =\frac{1}{N_\mathrm{s}-1} \sum_{n=1}^{N_\mathrm{s}} \left[ P_n(k_i)-\bar{P}(k_i) \right]\left[ P_n(k_j)-\bar{P}(k_j)\right],
\end{align}
where $P_n$ is the galaxy-velocity dipole power spectrum of the $n^{\rm th}$ simulated catalog obtained by running our pipeline on pairs of mock galaxy catalogs and simulated CMB maps, and $\bar{P}$ is the mean spectrum over mocks. We apply the Hartlap correction~\cite{Hartlap2007WhyYour} to the inverse of the covariance (or precision matrix) and de-bias our estimate through the change
\begin{align}
    \left[C^{-1}\right]_{\rm corrected} = \frac{N_\mathrm{s}-N_{\rm bins}-2}{N_\mathrm{s}-1} \; C^{-1}.
\end{align}
The final covariance is obtained by adding the noise-only and signal-only covariance estimates.

We use ten linearly-spaced bins in the range $k\in [0.0036, 0.048]\;h/$Mpc. The choice of the largest scale (smallest $k$) is driven by our ability to model wide-angle effects in the power spectrum. On scales larger than $k=2\pi/\chi(z_{\rm max})$, the large-angle corrections can no longer be modeled perturbatively~\cite{Benabou2024WideAngle}. For CMASS, this corresponds to $k\sim 0.0036\;h/$Mpc and we cut out data below this scale which masks our lowest $k$-bin so that our largest-scale bin is centered at $0.006 \;h/$Mpc. Our small-scale cut is chosen to avoid scale-dependent velocity bias. Ref.~\cite{Giri2022ExploringKSZ} finds mild scale-dependence in $N$-body simulations for scales above $0.05\;h/$Mpc. 

\begin{figure}
    \centering
    \includegraphics[width=\linewidth]{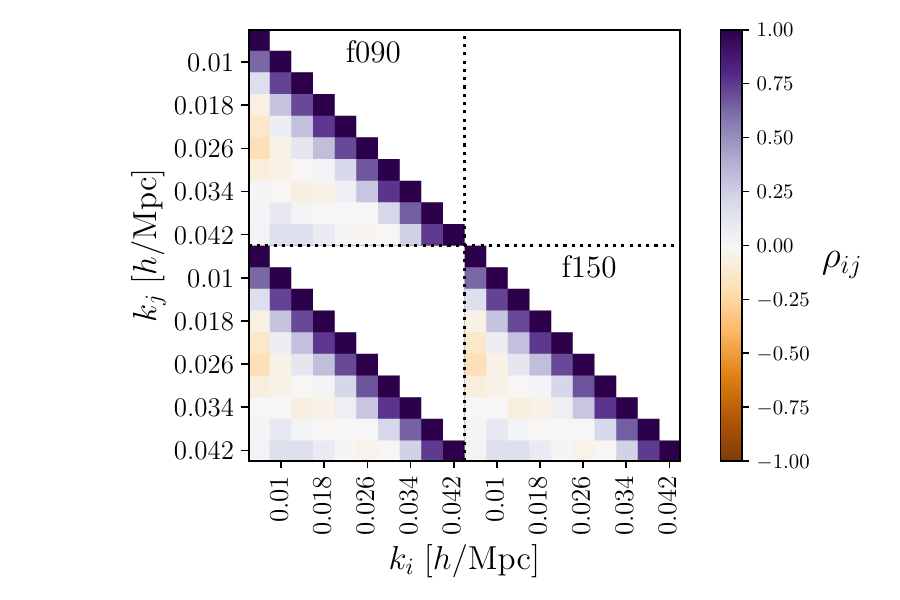}
    \caption{Correlation coefficient between galaxy-velocity dipoles obtained from the simulated f090 and f150 ACT maps and the NGC mock galaxy catalogs.}
    \label{fig:corr-coef}
\end{figure}

The correlation between the different $k$-bins and the two frequency maps is shown in Fig~\ref{fig:corr-coef}. We observe a high degree of correlation between the two frequency channels, as expected. 

\section{Null Tests}
We perform the reconstruction on the difference of the f090 and f150 maps. Since our reconstruction is linear in the temperature, we can simply subtract the f090 and f150 results at the power spectrum level. We define the null spectrum as $P^{g\hat{v}_r\;\rm null}_{\ell=1} \equiv P^{g\hat{v}_r\;\rm f090}_{\ell=1} - P^{g\hat{v}_r\;\rm f150}_{\ell=1}$. A non-zero null spectrum could indicate contamination from foregrounds or other frequency-dependent CMB secondaries. We list the corresponding probability-to-exceed (PTE) in the first row of Table~\ref{tab:null-ptes}.

\begin{figure}
    \centering
    \includegraphics[width=0.9\linewidth]{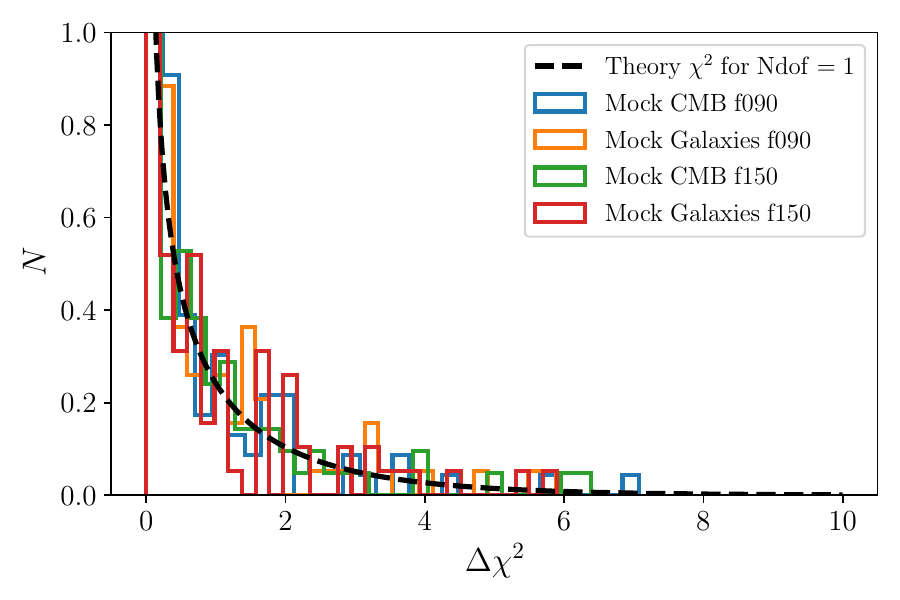}
    \caption{Normalized histogram of the change in $\chi^2$ between the best-fit $P_{gv}$ and the null hypothesis ($b_v=0$) for different combinations of data and mock observations. The distribution in $\Delta \chi^2$ follows a $\chi^2$ distribution with one degree of freedom according to KS tests.}
    \label{fig:null_hist}
\end{figure}

For our other null tests, we run our analysis with (1) the ACT map but substitute the galaxy catalog with a mock galaxy catalog from the QPM suite or with (2) the CMASS data and a mock ACT map. In both cases, we run the pipeline over the full set of 99 simulations. We perform this test on both frequencies and show the resulting $p$-values in Table~\ref{tab:null-ptes} with the full distribution of $\Delta \chi^2$ in Fig.~\ref{fig:null_hist}. The $p$-values are found using a Kolmogorov-Smirnov (KS) test the ensure the results of the null tests over the 99 simulations follow a $\chi^2$ distribution with one degree of freedom. We find that none of our null tests fail.

\begin{table}[]
    \centering
    \begin{tabular}{c | c | c | c }
        Frequency & Cap & Test  & PTE \\ \hline
        f090-f150 & NGC & $P_{gv}^{\rm null}$ & 0.926  \\
        f090 & NGC & Redshift shuffle & 0.240 \\
        f150 & NGC & Redshift shuffle & 0.131 \\
        f090 & NGC & Mock galaxies + true CMB &  0.177 \\
        f150 & NGC & Mock galaxies + true CMB &  0.246 \\
        f090 & NGC & True galaxies + mock CMB & 0.750  \\
        f150 & NGC & True galaxies + mock CMB & 0.674  \\ 
    \end{tabular}
    \caption{List of null test with corresponding PTEs or $p$-values (for KS tests).}
    \label{tab:null-ptes}
\end{table}

\section{Systematics\label{app:systematics}}
In this section, we outline the two main systematic effects we accounted for in our analysis. First is the biasing of the velocity amplitude by the thermal SZ effect. The second is the distortion at wide angles where the plane-parallel approximation used in our model for the $P_{gv}$ fails. 

\subsection{tSZ Bias}
While the quadratic estimator we use is robust against foreground contamination by symmetry considerations \cite{Smith2018KSZTomography}, the most massive galaxies in our sample can induce large scatter due to the thermal SZ effect, effectively appearing as a bias in the reconstruction~\cite{Schaan2016EvidenceFor,Schaan2021Combinedkinematic}. Since none of our null tests indicate the presence of tSZ bias, we do not mask any clusters in the our main analysis. We simply subtract the average of the reconstructed velocity to eliminate any offset caused by a constant tSZ signal. However, we do check for the impact of masking. To do this, (since the tSZ signal has such a strong mass dependence) we remove 910 galaxies that lie within 4 arcminutes of the center of clusters with mass above $M_{500c}>2.5\times 10^{14}\;M_\odot$ in the ACT DR5 cluster catalog~\cite{Hilton2021ACTClusters}. We find that the exact value of the mass threshold has little impact on the reconstructed velocity statistics. While the effect is small in the present work, tSZ mitigation may present a challenge for future surveys. For our analysis, the masking procedure we have tested does not lead to any significant changes in the constraints on $f_{\rm NL}$ or the value of the velocity bias. We observe a small reduction in SNR when applying a mask on more than $\sim 1000$ clusters.

\subsection{Window Function and Wide-Angle Terms}
From our use of a Yamamoto-like estimator, we have to include the survey window when fitting for $b_v$ and $f_{\rm NL}$~\cite{Benabou2024WideAngle}. To derive the window function multipoles we use the \texttt{smooth window} function in \texttt{pypower} on the galaxy random catalog. We compute up to multipole $\ell=8$ and order $n=1$ by evaluating~\cite{Beutler2019InterpretingMeasurements}
\begin{align}
    Q^{(n)}_\ell(s) \equiv(2\ell+1) \int d\Omega_s \int d^3 q \; W(\mathbf{q}) W(\mathbf{s}+\mathbf{q}) \mathcal{P}_\ell (\hat{\mathbf{s}}\cdot \hat{\mathbf{q}}),
\end{align}
where $W$ denotes the weights assigned at a specific location within the survey.

Given that our analysis focuses on large scales, we also have to account for wide-angle effects when modelling the galaxy-velocity dipole. At large angles, the choice of line-of-sight in the correlation function induces a coupling between odd and even multipoles and the window correlation function. Crucially, some of these corrections have a scale dependence resembling that of primordial non-Gaussianity and cannot be neglected~\cite{Benabou2024WideAngle}. Following the approach of Refs~\cite{Reimberg2016RedshiftSpace,Castorina2018BeyondThe}, we expand the two-point correlation function $\xi(s, d, \mu)$ around the point $s/d \ll 1$ giving 
\begin{align}
    \xi(s, d, \mu) = \sum_{n, \ell} \left(\frac{s}{d}\right)^n \xi^{(n)}_\ell(s) \mathcal{P}_\ell(\mu).
\end{align}
The higher order correlation function can be obtained directly from our zeroth order model for $P_\ell$ through
\begin{align}
    \xi_\ell^{(n)}(s) = \int\frac{k^2 dk}{2\pi^2} k^{-n} P_\ell(k) j_\ell(ks), \label{eq:xi_n_order}
\end{align}
where $j_\ell$ is the spherical Bessel function of order $\ell$. This standard calculation has been shown to introduce non-zero odd multipoles for the galaxy-galaxy power spectrum at first order. (At zeroth order, only the even multipoles are present due to the anisotropic galaxy-galaxy power spectrum being an even function of $\mu$.) In the case of the galaxy-velocity power multipoles, the converse happens: even multipoles only arise at higher order. This implies that we cannot use the same wide-angle expressions used previously with large-scale structure surveys (see e.g.~\cite{Beutler2019InterpretingMeasurements, Benabou2024WideAngle}). Note that it is customary to express Eq.~(\ref{eq:xi_n_order}) in terms of an integral over the linear matter power spectrum. However, as pointed out in Ref~\cite{Benabou2024WideAngle}, this is only valid if the galaxy bias is scale-independent. In the presence of primordial non-Gaussianity, the bias must be kept inside the integral and accounted for when computing the higher-order correlation functions.

We begin with our model for the anisotropic galaxy-velocity power spectrum (potentially including non-Gaussianity) which reads
\begin{align}
    P_{gv}(k, \mu) = \left(b+f\mu^2\right)\left(i\mu\frac{faH}{k}\right)P_{mm}(k).
\end{align}
The convolved, wide-angle-corrected multipoles are then obtained by convolution with the window and higher-order correlation functions using
\begin{align}
    P_\ell^{\rm wa}(k) = (-i)^\ell &(2\ell+1) \begin{pmatrix}
        \ell^\prime & \ell^{\prime\prime} & \ell \\
        0 & 0 & 0 
    \end{pmatrix} ^2
    \times \nonumber \\ & \sum_{\ell^\prime, \ell^{\prime\prime}} 
    \int ds \; \mathcal{M}_{\ell^\prime \ell^{\prime\prime}}(s) j_{\ell}(ks),
\end{align}
where the prefactor includes the Wigner-3j symbol and $\mathcal{M}_{\ell^\prime \ell^{\prime\prime}}$ is defined as the sum over higher order corrections such that
\begin{align}
    \mathcal{M}_{\ell^\prime \ell^{\prime\prime}}(s) \equiv  \sum_n \xi^{(n)}_{\ell^\prime}(s) Q^{(n)}_{\ell^{\prime\prime}}(s).
\end{align}
We make use of the FFTLog method implemented in the python package \texttt{mcfit}~\footnote{\url{https://github.com/eelregit/mcfit}} in order to evaluate the above integrals with log-spaced sample points efficiently. The final power spectrum including wide-angle and window selection effects is what is compared to observations.

\bibliography{bibliography}

\end{document}